\documentclass[prl,twocolumn,showpacs,preprintnumbers,superscriptaddress]{revtex4}

\usepackage{amsmath}
\usepackage{amssymb}
\usepackage{graphicx}
\usepackage{dcolumn}
\usepackage{bm}
\usepackage{epstopdf}

\begin{document}

\preprint{APS/123-QED}

\title{Fractal Dimension and Localization of DNA Knots}

\author{Erika Ercolini}
\affiliation{Laboratory of Physics of Living Matter, IPMC, Ecole Polytechnique
F\'{e}d\'{e}rale de Lausanne (EPFL), CH-1015 Lausanne, Switzerland}
\author{Francesco Valle$^*$}
\affiliation{Laboratory of Physics of Living Matter, IPMC, Ecole Polytechnique
F\'{e}d\'{e}rale de Lausanne (EPFL), CH-1015 Lausanne, Switzerland}
\author{Jozef Adamcik}
\affiliation{Laboratory of Physics of Living Matter, IPMC, Ecole Polytechnique
F\'{e}d\'{e}rale de Lausanne (EPFL), CH-1015 Lausanne, Switzerland}
\author{Guillaume Witz}
\affiliation{Laboratory of Physics of Living Matter, IPMC, Ecole Polytechnique
F\'{e}d\'{e}rale de Lausanne (EPFL), CH-1015 Lausanne, Switzerland}
\author{Ralf Metzler}
\affiliation{Nordic Institute for Theoretical Physics (NORDITA), Blegdamsvej
17, 2100 K{\o}benhavn \O, Denmark}
\author{Paolo De Los Rios}
\affiliation{Laboratory of Statistical Biophysics, ITP, Ecole Polytechnique
F\'{e}d\'{e}rale de Lausanne (EPFL), CH-1015 Lausanne, Switzerland}
\author{Joaquim Roca}
\affiliation{Instituto de Biologia Molecular de Barcelona, CID - CSIC, Jordi Girona  18 - 26, 08034  Barcelona, Spain}
\author{Giovanni Dietler}
\affiliation{Laboratory of Physics of Living Matter, IPMC, Ecole Polytechnique
F\'{e}d\'{e}rale de Lausanne (EPFL), CH-1015 Lausanne, Switzerland}

\date{\today}

\begin{abstract}
The scaling properties of DNA knots of different complexities were studied by atomic force microscope. Following two different protocols DNA knots are adsorbed onto a mica surface
in regimes of (i) strong binding, that induces a kinetic trapping of
the three-dimensional (3D) configuration, and of (ii) weak binding,
that permits (partial) relaxation on the surface. In (i) the gyration
radius of the adsorbed DNA knot scales with the 3D Flory exponent
$\nu\approx 0.58$ within error. In (ii), we find $\nu\approx 0.66$,
a value between the 3D and 2D ($\nu=3/4$) exponents, indicating an
incomplete 2D relaxation or a different polymer universality class. Compelling evidence is also presented for the localization of the knot crossings in 2D.
\end{abstract}


\maketitle

The first systematic study of knots was undertaken by Tait in
the 19th century \cite{tait}, following Kelvin's theory of vortex
atoms \cite{thomson}. During the 20th century progress was made
understanding knots in a topological framework and invariants were found to classify them  \cite{kauffmann,adams}. Experimentally,
knots remained elusive and difficult to study, but the discovery of their role in biological processes \cite{Wasserman,Krasnow} revived the interest in their properties. For example, knots on DNA  inhibit its separation
into single strands during replication, impede access to the full
genetic code during transcription, are implicated in gene regulation  \cite{pollock}, and influence DNA stability \cite{deibler}. Replication and transcription of
circular DNA are controlled by topoisomerases \cite{Wasserman} promoting questions on the detailed mechanism of knot detection \cite{Yan}. Finally, knots have also been found
in proteins in their native states \cite{Taylor}. The physical interest in the behavior of
DNA knots concerns two main questions: (i) the scaling properties of the radius of gyration $R_g$\cite{degennes1} and (ii) knot localization.

(i) From simulations and
scaling arguments, it is commonly accepted that the gyration radius of knots
to leading order scales as $R_g\simeq A\mathcal{L}^{\nu}$, for all knot types, as long
as the polymer is sufficiently long \cite{localize,Grosberg}, where $\mathcal{L}$ is the contour length. Here, we
quantify the Flory exponent $\nu$ of 3D and 2D configurations by determining the fractal dimension
of single DNA knots.

(ii) From a polymer physics interest, and to
understand better the action of topoisomerases and the
physiological role of DNA knots, it is crucial to find out whether
knot segregate into simply connected rings, with all essential crossings
confined in a knot region of contour length $s$ much smaller than
the overall chain length $\mathcal{L}$. 
Such localization has been predicted theoretically
as a consequence of entropic maximization \cite{Metzler}. 
Simulations in 3D yield a size distribution 
of the knot region that is peaked well below $\mathcal{L}$ for fixed knot types
\cite{Katritch}, and the size $s$ of the knot region scales as $s\sim \mathcal{L}^t$, with $t<1$ \cite{localize,Grosberg}. It is experimentally difficult to probe the predicted 
scaling behavior $s\sim \mathcal{L}^t$, since $\mathcal{L}$ would have 
to be varied significantly. This is at present out of reach given the available techniques used
to prepare the DNA knots.

Here we study the scaling properties and chain configuration of DNA knots
adsorbed onto a mica surface by atomic force microscope (AFM). Under strong
trapping conditions, we find that the gyration radius $R_g$ scales with 
the 3D exponent $\nu\approx 0.58$, while for weak trapping a larger value
is observed, $\nu\approx 0.66$. 
Moreover, from the analysis of single chain configurations 
we conclude that simple knots localize, as predicted from simulations studies.

Usually, knotted DNA obtained by topoisomerases is
studied by electron microscopy (EM) \cite{Liu, StasiakRecA}. For sufficient
contrast at the crossings, EM imaging requires coating of DNA by the protein
RecA, causing pronounced changes of physical parameters of DNA such as
stiffness and apparent diameter \cite{Wasserman,Krasnow,StasiakRecA}. As we document here, AFM can provide high
resolution images of bare DNA, permitting to probe
its unmasked polymeric properties. AFM has been applied to the visualization
of catenanes, resolving the crossings without protein coating \cite{Yamaguchi}.
Being a surface technique probing the 2D properties of the
adsorbed DNA, AFM does not give direct access to the 3D
conformation. However, we have shown that such information can indeed be extracted
from AFM images \cite{Valle}. In particular, we obtain the scaling exponent $\nu$ by determining the (average) fractal
dimension $d_f=1/\nu$ of individual DNA knot configurations \cite{falconer}. The configuration is found
at single molecule level from AFM images of DNA knots adsorbed onto a flat
surface out of solution and imaged by AFM in air. We consider two cases: (i) strong adsorption of
DNA knots on 3-aminopropyltriethoxy silane (APTES) modified mica; this strong kinetic trapping roughly corresponds
to a projection of the DNA knot onto the APTES-mica surface. As the fractal
dimension of a non-compact polymer in 3D fulfills $d_f\le 2$, its projection
onto a 2D surface preserves the value of $d_f$  \cite{falconer}, and we can indeed infer the
scaling exponent of the 3D structure. (ii) The second case is weak adsorption
onto untreated mica in the presence of Mg$^{2+}$ ions in solution. The ions
act as bridges between the negative mica surface and DNA charges. In this
case, the adsorption process allows for (partial) 2D relaxation of the knot
configuration.

Knotted and unknotted DNA was isolated from P4 phage capsids according to the protocols
given in \cite{Isaksen,Arsuaga}. All DNA knots are 10,346 base pairs (bp)
long, corresponding to a total contour length of $\approx 3.5\ \mu$m
for all knots types. The solutions contained a mixture of knot
types with a minimal crossing number ranging up to 30-40, with mean
complexity close to 30 minimal crossings. Further
extraction by electroelution from agarose gels was performed to yield
solutions containing DNA knots of low minimal crossing numbers in a
range from 3 to 6 \cite{Vologodskii,Adamcik}. We cannot exclude that
among these simple knots some could be unknots. The DNA knots were
free from disturbing supercoiling since they were obtained by joining
complementary ends without ligation of the strands ({\it nicked} DNA). 
Knotted DNA was
diluted in a buffer solution of 10 mM Tris-HCl, pH 7.6 and stored
at $4^{\circ}$C. For strong adsorption experiments \cite{Valle}, the
freshly cleaved mica substrate was positively charged by exposing
it to 3-aminopropyltriethoxy silane (APTES) vapors during 2 hours at
room temperature in a dry atmosphere \cite{Lyubchenko}. A 10 $\mu$l
drop of a knotted DNA solution was deposited onto the substrate
surface during 10 minutes and then rinsed with ultra-pure water (USF,
Elga). For the weak adsorption experiments, 10 $\mu$l DNA knot solution
(in TE buffer containing 5 mM MgCl$_2$) was deposited on freshly
cleaved mica during 10 minutes and then the sample was rinsed with
ultrapure water (see above). The samples were finally blown dry with
air. The DNA images were recorded by means of an AFM operated in
intermittent-contact mode, in order to reduce the effect
of lateral forces during scanning of the surface \cite{Garcia}. For
the case of strong adsorption, we checked that the sample remains
stable for weeks if kept in dry atmosphere and that upon imaging in
liquid the molecules do not rearrange, proving the irreversibility of
the adsorption \cite{Valle}. 

\begin{figure}
\includegraphics[width=8cm]{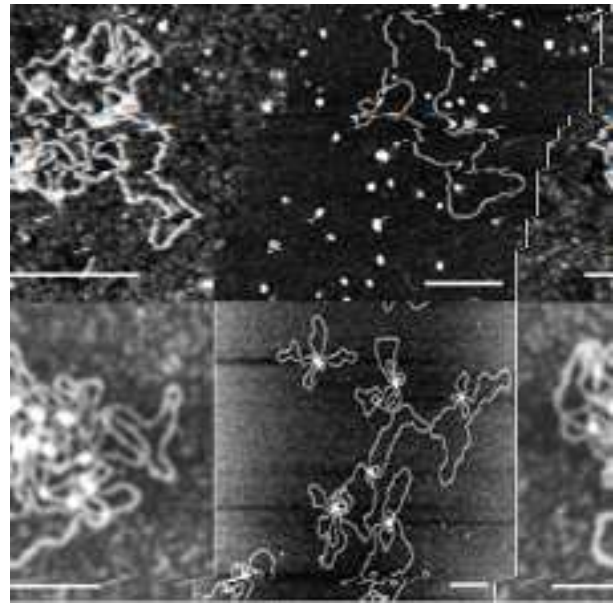}
\caption{AFM images of DNA knots selected from hundreds recorded. Strong adsorption case in the left column, weak adsorption on
the right. Top row shows simple knots ($\le$ 6 essential crossings),
bottom row complex knots (up to 30 essential crossings). Although statistically all knots behave similarly, single knots images clearly differ from each other. The scale bar
represents 250 nm on all images.}
\label{fig_01}
\end{figure}

Figure \ref{fig_01} depicts typical images of four knotted DNA molecules,
where the left column is obtained under strong adsorption conditions, and
the right column under weak adsorption. In the top row, knots with few
essential crossings ($\le$ 6) are shown, while the bottom row features
more complex knots with up to 30 essential crossings. It
is evident from these images that strong adsorption yields molecules
with many crossings, although
most of them are non-essential, while for the
weak adsorption likely only the essential crossings are present. The
latter case would indicate that the molecules are relaxed in a quasi
2D state.

\begin{figure}
\includegraphics[width=8cm]{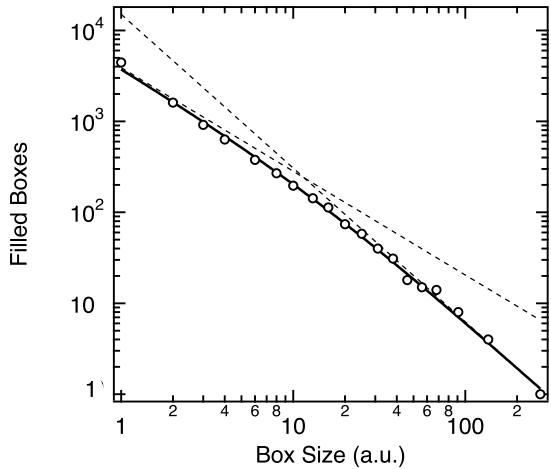}
\caption{Log-log plot of the number $N(L)$ of boxes of size $L$ filled with the knot versus $L$. The line is a fit to Eq.~(\ref{eq_02}).}
\label{fig_02}
\end{figure}

The fractal dimension was determined from the images using the box counting algorithm calculating the number of boxes $N(L)$ containing a part of the molecule as a function
of the box size $L$ \cite{falconer}. Each image of
a DNA knot was put through the algorithm and the resulting function was fitted
with Eq.~\ref{eq_02}. The curves for $N(L)$ present two scaling regimes, as shown in Fig.~\ref{fig_02}, similarly to the linear DNA case \cite{Valle}. On length scales
smaller than the crossover length $\ell_p$, DNA appears like a rigid rod with $d_1\approx 1$, while on scales larger than $\ell_p$, DNA appears flexible, and the relevant scaling exponent can be observed.
We therefore interpret $\ell_p$ as the persistence length of DNA (for double-stranded DNA under normal conditions $\ell_p\approx 45$ nm). The fractal dimension $d_f$ for both large and small length scales, and the persistence length $\ell_p$ from each individual molecule were
then averaged to yield the overall quantities $\bar{d_f}$ and
$\bar{\ell_p}$.

To describe the crossover of $N(L)$ from the initial rigid rod behavior to SAW behavior at $L=\ell_p$ we used the function

\begin{equation}
N(L)  =  a\left(\frac{L}{\ell_p}\right)^{-d_1}\left(1+\frac{L}{\ell_p}
\right)^{-(d_f-d_1)}\label{eq_02}
\end{equation}

Tab.~\ref{ExponentTable1} summarizes the results for the fractal
dimension and critical exponents for strongly adsorbed
DNA knots, representing averages over 50 DNA knots. The errors are taken as
the standard deviations for each case. Simple knots correspond to an essential
crossing number $\le 6$, while the complex knots comprise all
knots in a wide distribution of essential crossings with the most 
probable number of essential crossings around 30 \cite{Arsuaga}. 
In Tab.~\ref{ExponentTable1}, the first line contains the values 
for the unknots, that were purified from the same gel as the knots 
by taking the slowest band. Similarly to the knots, this circular DNA 
is not supercoiled because it is nicked. 
The results provide the first experimental proof that circular and 
knotted DNA under strong adsorption conditions correspond to a 
geometrical projection of the 3D configuration, and that the fractal 
dimension $d_f\approx 1.7$ is preserved upon strong adsorption, 
consistent with the previous findings for linear DNA \cite{Valle}. 
In particular, these results confirm the theoretical and numerical 
findings that the gyration radius of circular and knotted flexible 
polymers scales like $R_g\sim \mathcal{L}^{\nu}$ with $\nu\approx 0.588$ 
\cite{localize, Grosberg, Dobay}.

\begin{table}
\caption{Fractal dimension $d_f$, Flory exponent $\nu$, and persistence
length $\ell_p$ for the strong adsorption case.}
\begin{tabular}{ccc} \hline\hline \multicolumn{3}{c}{Strong
adsorption}\\
 \hline
  & $d_f$ & $\nu=1/d_f$\\ \hline Unknots & $1.711\pm
  0.042$ & $0.585\pm 0.014$\\  Simple knots & $1.685\pm 0.055$
  & $0.594\pm 0.019$\\
 Complex knots & $1.835\pm 0.076$ & $0.545\pm 0.024$\\ \hline\hline
\end{tabular}

\label{ExponentTable1}
\end{table}

Tab.~\ref{ExponentTable2} contains the results for the case when DNA
knots were deposited in presence of Mg$^{2+}$. The images in the right
column of Figure \ref{fig_01} clearly indicate that there
are significantly less crossings than in the strong adsorption 
case (left column). One
expects exponents that clearly differ from the 3D values, since relaxation
should lead to conformations closer to the ones of 2D polymers, characterized by
$\nu=0.75$. 
Yet, the scaling exponents for all three cases are 
significantly smaller, 
$\nu=0.66(1) < 0.75$. This may indicate that only a partial 2D relaxation 
takes place under weak adsorption conditions, or, that we are in presence of a different universality class. These points need to be further clarified using much longer DNA molecules.

\begin{table}
\caption{Same parameters as in Tab. \ref{ExponentTable1} but for the weak adsorption case.}
\begin{tabular}{ccc} \hline\hline \multicolumn{3}{c}{Weak
adsorption}\\
 \hline
  & $d_f$ & $\nu=1/d_f$\\ \hline Unknots & $1.491\pm
  0.037$ & $0.670\pm 0.017$\\ Simple knots & $1.530\pm 0.065$
  & $0.654\pm 0.028$\\
 Complex knots & $1.541\pm 0.086$ & $0.650\pm 0.036$\\ \hline\hline
\end{tabular}

\label{ExponentTable2}
\end{table}

In the analysis each
knot was treated separately and data were fitted with Eq.~(\ref{eq_02}). The results presented in Tables \ref{ExponentTable1} and \ref{ExponentTable2} are the averages of the
fractal dimensions or of the corresponding critical exponents. This procedure avoids the problem of knowing exactly the knot type and to average
over only one type of knot \cite{Dobay}. All values for $d_1$ were within $1.0\pm 0.1$ corroborating the stiff rod behavior on lengths scales shorter than $\ell_p$.

\begin{figure}
\includegraphics[width=7cm]{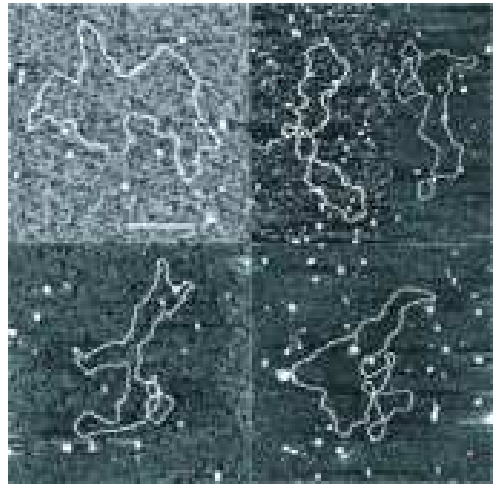}
\caption{Upper left : image of an unknot. The other images are DNA knots with a number of essential crossings smaller than 6 as separated from the agarose gels. Scale bar 250 nm.}
\label{fig_03}
\end{figure}

The average persistence length $\ell_p$ gained from the fits with (\ref{eq_02}) was in the range $7-10\ nm$ for strong adsorption (APTES-mica) and between $30-50$ nm for weak adsorption (Mg$^{2+}$). The small values in the first case are probably due to the very many unessential crossings, making the apparent persistence length determined from Eq. (\ref{eq_02}) much shorter. In fact, if we analyze the traceable parts of the DNA molecules  and apply the well known relation $\left<\cos(\theta(L+L_o)-\theta(L_o))\right>=e^{-L/\ell_p}$, where $\theta(L)$ is the direction of the tangent to the curve at $L$, the values were in range $30-50\ nm$ for strong and weak adsorption conditions, which is in agreement with the literature values\cite{Valle,Bednar,revet1,rivetti1}.

Apart from the critical exponents and the persistence length, from the AFM
images of weakly adsorbed simple knots we can deduce the localization behavior,
as (almost) all non-essential crossings are removed during the slow trapping
process. In Fig.~\ref{fig_03} we present additional images of DNA knots
deposited under weak adsorption conditions, illustrating the localization of
the essential crossings within a small region of the chain when allowed to
(partially) relax in 2D.
The upper left image is an unknot of the same length as the
knots and was extracted by electroelution from the first band of the agarose
gels. Almost all unknots we have imaged had no crossings, such that the
crossings on the other images, with significance, must be due to essential
crossings. Such localization into a comparatively small knot region was predicted for
2D self-avoiding chains in Ref.~\cite{Metzler}. This
localization is significant, and therefore the polymer phase is different
from dense or $\Theta$ conditions, for which delocalization was predicted
\cite{Hanke}. 

While future studies
with more advanced techniques of knot preparation are necessary to obtain
more detailed information, we demonstrated that AFM imaging and determination of the
fractal dimension of the chain configuration provide an outstanding way to analyze the behavior of DNA knots.

We thank A. Stasiak, J.B. Schvartzman, C. Vanderzande for fruitful
discussions. We acknowledge the support by the Swiss National Science
Foundation through the grants 200021-101851, and the Spanish Plan Nacional de I + D + I.

${}^{*}$Present Address: F. V., Department of Biochemistry "G. Moruzzi", University of Bologna, Via Irnerio 48, 40126 Bologna, Italy.

\end{document}